\documentclass[11pt]{article}
\usepackage{amsmath,euscript,amssymb}
\usepackage[T2A]{fontenc}
\usepackage[cp866]{inputenc}
\usepackage[dvips]{graphicx}
\def\v1{\vspace{1cm}}
\def\be{\begin{equation}}
\def\ee{\end{equation}}
\def\bc{\begin{center}}
\def\ec{\end{center}}

\def\vh{\varphi}

\newcommand{\bea}{\begin{eqnarray}}
\newcommand{\eea}{\end{eqnarray}}

\begin{document}
\title{Quantum Gravity as Theory of ``Superfluidity''}
\author{ B.M. Barbashov${}^{1}$,  V.N. Pervushin${}^{1}$, A.F. Zakharov${}^{1,2}$, and
V.A. Zinchuk${}^{1}$,
 \\[0.3cm]
{\normalsize\it 1. Joint Institute for Nuclear Research},\\
 {\normalsize\it 141980, Dubna, Russia.} \\
 {\normalsize\it 2. National Astronomical Observatories of Chinese Academy
of Sciences},\\
 {\normalsize\it  Beijing 100012, China} }

\date{\empty}
\maketitle
\medskip

\begin{abstract}
 {\noindent

 A version of the cosmological perturbation theory in general
 relativity (GR)
 is developed,
 where the cosmological scale factor is identified with
 spatial averaging of the metric determinant logarithm
  and the cosmic evolution acquires the pattern of a superfluid
  motion: the absence of ``friction-type'' interaction,
  the London-type wave function, and the Bogoliubov condensation of
  quantum universes.
 This identification keeps the number of variables of GR and
  leads to a new type of  potential perturbations. A set of
  arguments is given in  favor of that this ``superfluid'' version
  of GR is in agreement with the observational data.
}
\end{abstract}

\vspace{1cm}


\begin{center}
{
  ``Long Communication'' at the  XXVIII  Spanish
Relativity Meeting

``A Century of Relativity Physics''

\vspace{.1cm}

 Oviedo, Spain, 6-10 Sept. 2005

 http://fisi24.ciencias.uniovi.es/ere05.html

\vspace{.1cm}

To be published in ``AIP Conference Proceedings''}
\end{center}

\vspace{1cm}

\tableofcontents

\newpage
\section{Introduction}

Separation of the cosmological scale factor from metrics in
 general relativity (GR)
 is well known as the cosmological perturbation theory proposed by
 Lifshitz \cite{Lif} and applied as a basic tool
 for analysis of modern observational data in astrophysics and cosmology
 \cite{bard}. However, the number of variables of the Lifshitz
 theory differs from GR.
 In the present paper,   a new version of
 the cosmological perturbation theory  \cite{origin} is discussed, where
 the  logarithm of cosmological scale factor coincides with
 spatial averaging the metric determinant logarithm, so that
 the number of variables of GR is conserved.
 The conservation of the number of variables
  allows us to solve the energy constraint,
 fulfil the Dirac Hamiltonian reduction and quantization.
 This Quantum Gravity
 has some attributes of the
 theory of superfluidity of the type of Landau's ``superfluid'' dynamics
 \cite{Lan},
  London's unique wave function \cite{Lon},
 and   Bogoliubov's squeezed condensate  \cite{B} and gives  us
  a new possibility  to
  explain the ``CMBR primordial power spectrum'' and other topical
 problems of modern cosmology.

\section{``Superfluid'' motion and potential perturbations}

     GR
    was given by Einstein and Hilbert 90 years ago
     with
    the  {\it``dynamic''} action
 \be\label{gr}
 S=\int d^4x\sqrt{-g}\left[-\frac{\vh_0^2}{6}R(g)
 +{\cal L}_{(\rm M)}\right],
 \ee
 where  $\varphi^2_0=\frac{3}{8\pi}M^2_{\rm Planck}$ is the Newton
constant,
 ${\cal L}_{(\rm M)}$  is the matter Lagrangian,
     and an
 {\it``interval''}
$
 g_{\mu\nu}dx^\mu dx^\nu\equiv\omega_{(\alpha)}\omega_{(\alpha)}=
 \omega_{(0)}\omega_{(0)}-
 \omega_{(1)}\omega_{(1)}-\omega_{(2)}\omega_{(2)}-\omega_{(3)}\omega_{(3)}
 $ 
 presented here
 by the components of an
orthogonal
 simplex of reference $\omega_{(\alpha)}$.

  These simplex components
  in the reference frame  of the Dirac -- ADM Hamiltonian
  approach to GR \cite{dir} take the form
 \be \label{adm}
 \omega_{(0)}=\psi^6N_{\rm d}dx^0,~~~~~~~~~~~
 \omega_{(b)}=\psi^2 {\bf e}_{(b)i}(dx^i+N^i dx^0);
 \ee
 here  ${\bf e}_{(a)i}$ are triads of the spatial metrics with $\det |{\bf
 e}|=1$, $N_{\rm d}$ is the Dirac lapse function, $N^i$ is  shift
 vector, and $\psi$ is a determinant of the spatial metric.
 All these components were considered by Dirac as the potentials
 except of two transverse triads distinguished by the constraints
 $\partial_j{\bf e}^j_{(a)}=0$ and the zero determinant momentum
 $p_\psi=0$. Latter  contradicts to Friedmann-type evolution
 in the homogeneous approximation of GR, where
  the determinant component is identified with the cosmological scale
  factor ($\psi^2\simeq a(x^0)$) and its momentum coincides with
  the Hubble parameter.
 The question appears about the status
  of the Hubble parameter in the Dirac Hamiltonian approach
  \cite{dir}.

  To answer this question we consider
 the cosmological perturbation theory
 \cite{Lif,bard}
 $g_{\mu\nu}=a^2(x^0)\widetilde{g}_{\mu\nu}$ in terms of
 the Dirac -- ADM variables, where one can get the new lapse
 function
 $\widetilde{N}_d=[\sqrt{-\widetilde{g}}~\widetilde{g}^{00}]^{-1}
 =a^{2}{N}_d$ and spatial determinant
 $\widetilde{\psi}=(\sqrt{a})^{-1}\psi
 $.

 After the separation of the spatial metric determinant
in  the
   curvature
 $
 \sqrt{-g}R(g)=a^2\sqrt{-\widetilde{g}}R(\widetilde{g})-6a
\partial_0\left[\widetilde{N}_d^{-1}{\partial_0a}\right]$
 the determinant part
 of the Lagrangian takes the form
 \be\label{sd1-0}
 L=-\vh^2_0\int d^3x\widetilde{N}_d
 \left[{\vphantom{\int}}4a^2~
 {{(\overline{v_{\psi}})}^2}+
 \left(\frac{\partial_0a}{\widetilde{N}_d}\right)^2\right]
 -\fbox{$~2\vh^2_0 {\partial_0a^2}\int\limits_{}^{}
d^3x\overline{v_{\psi}}$}+...,
 \ee
  where
 $ \overline{v_{\psi}}=\left[
 (\partial_0-N^l\partial_l)\log{\widetilde{\psi}}-
 \frac16\partial_lN^l\right]/{\widetilde{N}_d}$
 is a local velocity. One can see that this Lagrangian
 contains the velocity-velocity interaction distinguished by the
 box in Eq. (\ref{sd1-0}). This interaction mixes the canonical momenta
 \bea\label{sd1-1}
 P_a&\equiv&
  \frac{\partial {L}}{\partial(\partial_0 a)}=
 -\vh_0^2\int d^3x\left[  \fbox{$4a~\overline{v_\psi}$}~+
 2\frac{\partial_0a}{\widetilde{N}_d}\right],\\\label{sd1-2}
\overline{p_{\psi}}&=&\frac{\partial {\cal L}}{\partial(\partial_0
\log  \widetilde{\psi})}=-2\vh_0^2a\left[4a \overline{v_\psi}+
  \fbox{$2~\dfrac{\partial_0a}{\widetilde{N}_d}$}~\right],
 \eea
  so that the integral of the local momentum $\overline{p_\psi}$
  coincides with the  one of cosmic motion
 \be\label{sd1-3}
   \int d^3x \overline{p_{\psi}} =
 -2\vh_0^2a \int d^3x \left[4a \overline{v_\psi}+
  \fbox{$2~\dfrac{\partial_0a}{\widetilde{N}_d}$}~\right]
 \equiv   \fbox{$2a P_{a}$},
 \ee
  and the Hamiltonian approach is failure. The reason of that is the
  increase of the number of variables after the separation
  $a(x^0)$.

 In order to keep the number of variables, we identify $\log \sqrt{a}$ with
 the  spatial volume ``averaging'' of $\log{\psi}$:
 $
 \log \sqrt{a}=\langle \log{\psi}\rangle\equiv\int
 d^3x\log{\psi}/{V_0},
 $
 where $V_0=\int d^3x  < \infty$ is a finite
 volume. In this case, the new determinant variable $\widetilde{\psi}$
  and its velocity satisfy  the identities
 \be\label{non1}
 \int d^3x \log\widetilde{\psi}=\int d^3x \left[\log{\psi}
 -\left\langle{ \log{\psi}}\right\rangle\right]\equiv 0,
 ~~~~~~~~~~~~\int\limits_{}^{} d^3x\overline{v_{\psi}}\equiv 0.
 \ee
 This means that all the box terms in Eqs (\ref{sd1-0})-- (\ref{sd1-3})
disappear
 together with the ``friction-type'' velocity-velocity
 interaction. The momenta (\ref{sd1-1}) and  (\ref{sd1-2}) are
 completely separated
  \bea\label{sd1-11}
 P_a&\equiv&=
 -2V_0\vh_0^2\partial_0a
\langle(\widetilde{N}_d)^{-1}\rangle=-2V_0\vh_0^2\frac{da}{d\zeta}\equiv-2V_0\vh_0^2a',\\\label{sd1-21}
\overline{p_{\psi}}&=&\frac{\partial {\cal L}}{\partial(\partial_0
\log  \widetilde{\psi})}=-8\vh_0^2a^2 \overline{v_\psi},
 \eea
 here the averaging  $\langle(\widetilde{N}_d)^{-1}\rangle$ determines
 the  time interval
 $d\zeta={\langle(\widetilde{N}_d)^{-1}\rangle}^{-1}dx^0$
 that is invariant with respect to reparametrizations of the coordinate
 evolution parameter $x^0$. In this
 case,  the scale transformation $g_{\mu\nu}=a^2(x^0)\widetilde{g}_{\mu\nu}$
 converts the GR action (\ref{gr}) into the ``friction-free'' one
 \be\label{1gr}
 S[\vh_0]=\widetilde{S}[\vh]-
 \int dx^0 (\partial_0\vh)^2\int \frac{d^3x}{\widetilde{N}_d},
 \ee
 where $\widetilde{S}[\varphi]$
  is the action (\ref{gr})  in
 terms of metrics $\widetilde{g}$ and
 the running scale $\vh(x^0)=\vh_0a(x^0)$ of all masses including
 the Planck one $\vh_0$.
 The
 energy constraint ${\delta S[\vh_0]}/{\delta  \widetilde{N}_d}=0$ takes
 algebraic form \cite{pp} with respect to the
 invariant local lapse function
 $N_{\rm inv}\equiv{\widetilde{N}_d\langle(\widetilde{N}_d)^{-1} \rangle}$
 \be\label{nph}
 -
 \frac{\delta \widetilde{S}[\vh]}{\delta  \widetilde{N}_d}
 \equiv
\widetilde{T}^0_0=\left[\frac{\partial_0\varphi}{\widetilde{N}_d}\right]^2
 =\frac{1}{N_{\rm inv}^2}\left[\frac{d\vh}{d\zeta}\right]^2=
 \frac{\vh'^2}{N_{\rm inv}^2},
 \ee
 where $\widetilde{T}^0_0$ is the local energy density.
 This equation has the  resolution in both the local sector
\be\label{13ec}
 N_{\rm inv}\equiv{\widetilde{N}_d\langle(\widetilde{N}_d)^{-1} \rangle}
  ={{\left\langle\sqrt{{\widetilde{T}^0_0}}\right\rangle}}
  \left({\sqrt{{\widetilde{T}^0_0}}}\right)^{-1}
 \ee
 and the global one
\be\label{13c}
 \zeta(\varphi_0|\varphi)
 \equiv\int dx^0 \langle{(\widetilde{N}_d)^{-1}}\rangle^{-1}
 =\pm
 \int_{\vh}^{\vh_0}d\widetilde{\vh}
 {{\left\langle
 \sqrt{\widetilde{T}^0_0(\widetilde{\vh})}\right\rangle}}^{-1}
 \ee
 where  the invariant time $\zeta$ is connected with the scale
 factor by the Hubble-like law.

 The explicit dependence of $\widetilde{T}^0_0$ on $\widetilde{\psi}$
  was given by Lichnerowicz  \cite{L}
 \be\label{La-2}
 \widetilde{T}^0_0= \widetilde{\psi}^{7}\hat \triangle
 \widetilde{\psi}+
  \sum\limits_{I} \widetilde{\psi}^I a^{\frac{I}{2}-2}\tau_I,
  ~~~~~~\tau_I\equiv\langle\tau_I\rangle+\overline{\tau_I},
 \ee
where $\hat \triangle
 \widetilde{\psi}\equiv\dfrac{4\varphi^2}{3}\partial_{(b)}
 \partial_{(b)}\widetilde{\psi}$ is the
 Laplace operator and  $\tau_I$ is partial energy density
  marked by the index $I$ running a set of values
   $I=0$ (stiff), 4 (radiation), 6 (mass), 8 (curvature), 12
   ($\Lambda$-term)
 in correspondence with a type of matter field contributions.
 The negative contribution $-({16}/{\vh^2})\overline{p_{\psi}}^2$ of the
 spatial determinant momentum  in the energy
 density $\tau_{I=0}$
can be removed by the Dirac constraint \cite{dir} of the
 zero velocity of the spatial volume element
 \be\label{La-6}
 \overline{p_{\psi}}=-8\vh^2\frac{1}{\widetilde{N}_d}\left[
 (\partial_0-N^l\partial_l)\ln{\widetilde{\psi}}-\frac16\partial_lN^l\right]=
 -8\vh^2\frac{\partial_0\widetilde{\psi}^6-\partial_l
 [\widetilde{\psi}^6N^l]}{\widetilde{\psi}^6\widetilde{N}_d}=0.
 \ee
 In the class of functions $\overline{F}=F-\langle F\rangle$, the classical
 equation ${\delta S}/{\delta \log \widetilde{\psi}}=0$ takes the
 form
 \be\nonumber
 \widetilde{N_d}\widetilde{\psi}
 \frac{\partial \widetilde{T}_0^0}{\partial \widetilde{\psi}}+
 \widetilde{\psi}\triangle \left[
 \frac{\partial \widetilde{T}_0^0}{\partial
 \triangle\widetilde{\psi}}\widetilde{N_d}\right]=0 ~
 \to~
 7\overline{N_{\rm inv}\widetilde{\psi}^7\hat \triangle\widetilde{\psi}}
 \!+\!\overline{\widetilde{\psi}\hat \triangle[N_{\rm inv}\widetilde{\psi}^7]}
 \!+\!\sum\limits_{I} I\overline{\widetilde{\psi}^I a^{\frac{I}{2}-2}\tau_I}=0
 \ee
 and  gives both the determinant
 $\widetilde{\psi}=1+\overline{\mu}$
 and the lapse function (\ref{13ec})
 ${N_{\rm inv}}\widetilde{\psi}^{7}=1-\overline{\nu}$.

 The
 first order of this equation
 determines $\overline{\mu}$ and $\overline{\nu}$ in the form of a sum
 \cite{origin}
  \bea\label{2-17}
 {\overline{\mu}}&=&\frac{1}{2}\int d^3y\left[D_{(+)}(x,y)
\overline{T}_{(+)}^{(\mu)}(y)+
 D_{(-)}(x,y) \overline{T}^{(\mu)}_{(-)}(y)\right],\\\label{2-18}
 {\overline{\nu}}&=&\frac{1}{2}\int d^3y\left[D_{(+)}(x,y)
\overline{T}^{(\nu)}_{(+)}(y)+
 D_{(-)}(x,y) \overline{T}^{(\nu)}_{(-)}(y)\right],
  \eea
 where
 $$\beta=\sqrt{1+[\langle \tau_{(2)}\rangle-14\langle
\tau_{(1)}\rangle]/(98\langle \tau_{(0)}\rangle)},$$
 \be\label{cur1}\overline{T}^{(\mu)}_{(\pm)}=
 [1\pm 49\beta]\overline{\tau_{(0)}}\mp 7\beta\overline{\tau_{(1)}},
 ~~~~~~~
 \overline{T}^{(\nu)}_{(\pm)}=
 [7\pm(14\beta)^{-1}]\overline{\tau_{(0)}}-\overline{\tau_{(1)}}
 \ee
 are the local currents, $D_{(\pm)}(x,y)$ are the Green functions satisfying
 the equations
 \bea\label{2-19}
 [\pm \hat m^2_{(\pm)}-\hat \triangle
 ]D_{(\pm)}(x,y)=\delta^3(x-y),
 \eea
 where $\hat m^2_{(\pm)}= 14 (\beta\pm 1)\langle \tau_{(0)}\rangle \mp
\langle \tau_{(1)}\rangle$,
$\tau_{(n)}=\sum_II^na^{\frac{I}{2}-2}\tau_{I}$.
  In the case of point mass distribution in a finite volume $V_0$ with the
zero pressure
  and  the  density
  $\overline{\tau_{(0)}}(x)=\overline{\tau_{(1)}}(x)
/6\equiv  M\left[\delta^3(x-y)-{1}/{V_0}\right]$,
 solutions   (\ref{2-17}),  (\ref{2-18}) take
 a very interesting form
 \bea\label{2-21}
  \widetilde{\psi}&=1+\overline{\mu}(x)=1+
  \dfrac{r_{g}}{4r}\left[{\gamma_1}e^{-m_{(+)}(z)
 r}+ (1-\gamma_1)\cos{m_{(-)}(z)
 r}\right],\\\label{2-22}
 N_{\rm inv}\widetilde{\psi}^{7}&=
 1-\overline{\nu}(x)=1-
 \dfrac{r_{g}}{4r}\left[(1-\gamma_2)e^{-m_{(+)}(z)
 r}+ {\gamma_2}\cos{m_{(-)}(z)
 r}\right],
 \eea
 where
 $
  {\gamma_1}=\dfrac{1+7\beta}{2},~~~
 {\gamma_2}=\dfrac{14\beta-1}{28\beta},~~
 r_{g}=\dfrac{3M}{4\pi\vh^2},~~
 r=|x-y|.
 $
 The zero volume velocity (\ref{La-6})
 gives the diffeo-invariant shift of the coordinate
  origin
 \be \label{2-23}
\langle{(\widetilde{N}_d)^{-1}}\rangle{N}^i=\left(\frac{x^i}{r}\right)\left(\frac{\partial_\zeta
V}{\partial_r V}\right),~~~
 ~~~V(\zeta,r)=\int\limits_{}^{r}d\widetilde{r}
 ~\widetilde{r}^2\widetilde{\psi}^{6}(\zeta,\widetilde{r}).
  \ee
In the infinite volume limit $\langle \tau_{(n)}\rangle=0,~a=1$
 solutions (\ref{2-21}) and  (\ref{2-22}) coincide with
 the isotropic version of  the Schwarzschild solutions:
 $\widetilde{\psi}=1+\dfrac{r_g}{4r}$,~
 ${N_{\rm inv}}\widetilde{\psi}^{7}=1-\dfrac{r_g}{4r}$,~$N^k=0$.

 The main differences of the
 superfluid-type version of GR from the Lifshitz version \cite{Lif,bard} are the
 potential perturbations of the scalar components ${N_{\rm inv}},
 \widetilde{\psi}$
 instead of the kinetic ones and the
 nonzero shift vector $N^k\not =0$ determined by Eq. (\ref{La-6}).
  Recall that just the kinetic perturbations
   are responsible for the ``primordial power spectrum'' in
  the inflationary model \cite{bard}. The problem appears to
  describe CMBR by the potential perturbations given by Eqs. (\ref{2-17}) --
(\ref{2-23}).

 \section{London wave function \& Bogoluibov condensation}
 The main consequence of the separation of the cosmological  scale
 factor is the London-type globalization of the
  energy constraint (\ref{13ec}). It fixes only
  the scale momentum $P_\vh$
  \be\label{La-1}
  P_\vh^2\equiv [P_a/\vh_0]^2=E_\vh^2\equiv
  \left[2\int d^3x\sqrt{\widetilde{T}_0^0}~
  \right]^2,
  \ee
  in contrast to the energy constraint in GR without the cosmological  scale
 factor.
 The ``reduced'' action can be obtained as
   values of the Hamiltonian  action  for
 the energy constraint $P^2_{\varphi}=E^2_{\varphi}$ \cite{pp}:
 \be\label{2ha2} S[\vh_I|\vh_0]|_{P_\vh=\pm E_\vh} =
 \int\limits_{\vh_I}^{\vh_0}d\widetilde{\vh} \left\{\int d^3x
 \left[\sum\limits_{  F}P_{  F}\partial_\vh F
 +{C}\mp2\sqrt{\widetilde{T}_0^0(\widetilde{\vh})}\right]\right\},
\ee
 where ${C}$ is the sum of all Dirac constraints \cite{dir,pp}. Momenta
 ${P_\vh}_{\pm}=\pm E_\vh$
   become  the generator of evolution of all variables with respect to the
  evolution parameter $\vh$ \cite{pp} forward and backward,
 respectively. The negative energy problem can be solved by
 the primary quantization of the energy constraint
 $[{P^2_\vh}-E^2_\vh]\Psi_{\rm u}=0$ and the secondary quantization
 $\Psi_{\rm u}=(1/\sqrt{2E_\vh)}[A^++A^-]$ by the Bogoliubov
 transformation $ A^+=\alpha
 B^+\!+\!\beta^*B^-$, in order to diagonalize the equations of
 motion by the condensation of ``universes''
 $<0|\dfrac{i}{2}[A^+A^+-A^-A^-]|0>=R(\vh)$
 and describe  cosmological creation of a  ``number'' of universes
  $<0|A^+A^-|0>=N(\vh)$
  from the stable Bogoliubov vacuum  $B^-|0>=0$.
 Vacuum postulate $B^-|0>=0$ leads to an arrow of the invariant
time $\zeta\geq 0$ (\ref{13c}) and its absolute point of reference
$\zeta= 0$ at the moment of creation $\vh=\vh_I$ \cite{origin,pp};
whereas the Planck value of
 the running mass scale $\vh_0=\vh(\zeta=\zeta_0)$ belongs to the present
 day moment $\zeta_0$. The reduced action (\ref{2ha2}) shows us
 that the initial data at the beginning $\vh=\vh_I$ are independent of
  the present-day ones at  $\vh=\vh_0$,
  therefore
  the proposal about an existence of the  Planck epoch $\vh=\vh_0$
   at the beginning \cite{bard} looks
  very doubtful.

\section{W-,Z- Factory versus the Planck epoch}

  The low-energy
 expansion of the {\it``reduced action''} (\ref{2ha2}) over the
 field density ${T}_{{\rm sm}}$
 $$2 d\vh \sqrt{\widetilde{T}_0^0}= 2 d\vh
 \sqrt{\rho_{0}(\vh)+{T}_{{\rm sm}}}
 =
 d\vh
 \left[2\sqrt{\rho_0(\vh)}+
 {{T}_{{\rm sm}}}/{\sqrt{\rho_0(\vh)}}\right]+...$$
 gives the sum:
 $S^{(+)}|_{\rm
 constraint}= S^{(+)}_{\rm cosmic}+S^{(+)}_{\rm
 field}+\ldots$, where
 $S^{(+)}_{\rm cosmic}[\varphi_I|\varphi_0]= -
 2V_0\int\limits_{\vh_I}^{\vh_0}\!
 d\vh\!\sqrt{\rho_0(\vh)}$ is the reduced  cosmological action
 and
 \be\label{12h5} S^{(+)}_{\rm field}=
 \int\limits_{\eta_I}^{\eta_0} d\eta\int d^3x
 \left[\sum\limits_{ F}P_{ F}\partial_\eta F
 -{T}_{\rm sm }\right]
 \ee
 is the standard field action
 in terms of the conformal time:
 $d\zeta=d\eta=d\vh/\sqrt{\rho_0(\vh)}$
 in the conformal flat space--time with running masses
 $m(\eta)=a(\eta)m_0$.

 This low-energy expansion
 identifies the ``conformal quantities''
  with the observable ones including the conformal time $d\eta$,
  instead of $dt=a(\eta)d\eta$, the coordinate
 distance $r$, instead of Friedmann one $R=a(\eta)r$, and the conformal
 temperature $T_c=Ta(\eta)$, instead of the standard one $T$.
 In this case,
 the
 cosmological  redshift of the spectral lines 
 $$
\frac{E_{\rm emission}}{E_0}=\frac{m_{\rm atom}(\eta_0-r)}{m_{\rm
atom}(\eta_0)}\equiv\frac{\vh(\eta_0-r)}{\vh_0}=a(\eta_0-r)
=\frac{1}{1+z}
$$
is explained by the running masses $m=a(\eta)m_0$ in action
(\ref{12h5}) \cite{Narlikar,039,Danilo}.

The conformal observable distance  $r$ loses the factor $a$, in
comparison with the nonconformal one $R=ar$. In the units of
``conformal
 quantities'' the Supernova data \cite{snov,SN} are
consistent with the dominance of the stiff (rigid) state,
$\Omega_{\rm Rigid}\simeq 0.85 \pm 0.15$, $\Omega_{\rm
Matter}=0.15 \pm 0.15$ \cite{039,Danilo}. If $\Omega_{\rm
Rigid}=1$, we have the square root dependence of the scale factor
on conformal time $a(\eta)=\sqrt{1+2H_0(\eta-\eta_0)}$. Just this
time dependence of the scale factor on
 the measurable time (here -- conformal one) is used for description of
 the primordial nucleosynthesis \cite{three}.
 This stiff state is formed by a free scalar field
 when $E_\vh=2V_0\sqrt{\rho_0}=Q/\vh$ and $Q/2V_0=H_0\vh^2_0=H_I\vh^2_I$
 is an integral of motion.


 These initial data $\vh_I$ and $H_I$ can be determined by the
 parameters of matter cosmologically created from the
 vacuum  at the beginning of a universe $\eta\simeq 0$.

 The Standard
 Model (SM) density ${T}_{{\rm sm}}$ in action (\ref{12h5})
  shows
 us that W-, Z- vector bosons have maximal probability of this
 cosmological creation
 due to their mass singularity~\cite{114:a}. One can introduce the notion of
 a particle in a universe if the Compton length of a particle
 defined by its inverse mass
 $M^{-1}_{\rm I}=(a_{\rm I} M_{\rm W})^{-1}$ is less than the
 universe horizon defined by the inverse Hubble parameter
 $H_{\rm I}^{-1}=a^2_{\rm I} (H_{0})^{-1}$ in the
 stiff state. Equating these quantities $M_{\rm I}=H_{\rm I}$
 one can estimate the initial data of the scale factor
 $a_{\rm I}^2=(H_0/M_{\rm W})^{2/3}=10^{-29}$ and the primordial Hubble
parameter
 $H_{\rm I}=10^{29}H_0\sim 1~{\rm mm}^{-1}\sim 3 K$.
 Just at this moment there is  an effect of intensive
  cosmological creation of the vector bosons described in \cite{114:a};
 in particular, the distribution functions of the longitudinal   vector bosons
demonstrate us a large contribution of relativistic momenta.
 Their conformal (i.e. observable) temperature $T_c$
 (appearing as  a consequence of
 collision and scattering of these bosons) can be estimated
from the equation in the kinetic theory for the time of
establishment of this temperature $ \eta^{-1}_{\rm relaxation}\sim
n(T_c)\times \sigma \sim H $, where $n(T_c)\sim T_c^3$ and $\sigma
\sim 1/M^2$ is the cross-section. This kinetic equation and values
of the initial data $M_{\rm I} = H_{\rm I}$ give the temperature
of relativistic bosons \be\label{t}
 T_c\sim (M_{\rm I}^2H_{\rm I})^{1/3}=(M_0^2H_0)^{1/3}\sim 3 K
\ee as a conserved number of cosmic evolution compatible with the
Supernova data \cite{039,snov,SN}.
 We can see that
this  value is surprisingly close to the observed temperature of
the CMB radiation
 $ T_c=T_{\rm CMB}= 2.73~{\rm K}$.

 The primordial mesons before
 their decays polarize the Dirac fermion vacuum
 (as the origin of axial anomaly)
 and give the
 baryon asymmetry frozen by the CP-violation.
The
 value of the baryon--antibaryon asymmetry
of the universe following from this axial anomaly was estimated in
\cite{114:a} in terms of the coupling constant of the
superweak-interaction
 \be\label{a}
 n_b/n_\gamma\sim X_{CP}= 10^{-9}.
 \ee The boson
life-times     $\tau_W=2H_I\eta_W\simeq
\left({2}/{\alpha_W}\right)^{2/3}\simeq 16,~ \tau_Z\sim
2^{2/3}\tau_W\sim 25
 $ determine the present-day visible
baryon density
\be\label{b}\Omega_b\sim\alpha_W=\alpha_{QED}/\sin^2\theta_W\sim0.03.\ee
All these results (\ref{t}) -- (\ref{b})
 testify to that all  visible matter can be a product of
 decays of primordial bosons, and the observational data on CMBR
 can reflect  parameters of the primordial bosons, but not the
 matter at the time of recombination. In particular,
 the length of  the semi-circle on the surface of  the last emission of
photons at the life-time
  of W-bosons
  in terms of the length of an emitter
 (i.e.
 $M^{-1}_W(\eta_L)=(\alpha_W/2)^{1/3}(T_c)^{-1}$) is
 $\pi \cdot 2/\alpha_W$.
 It is close to $l_{min}\sim  210 $ of CMBR,
 whereas $(\bigtriangleup T/T)$ is proportional to the inverse number of
emitters~
 $(\alpha_W)^3 \sim    10^{-5}$. 

 The temperature history of the expanding universe
 copied in the ``conformal quantities'' looks like the
 history of evolution of masses of elementary particles in the cold
 universe with the conformal temperature $T_c=a(\eta)T$
 of the cosmic microwave background.
  In the superfluid version of cosmology \cite{039}, the
  dominance of stiff state $\Omega_{\rm Stiff}\sim 1$ determines the parameter
  of spatial oscillations
  $\hat m^2_{(-)}=\frac{6}{7}H_0^2[\Omega_{\rm R}(z+1)^2+\frac{9}{2}\Omega_{\rm
  Mass}(z+1)]$. The values of red shift in the recombination
  epoch $z_r\sim 1100$ and the clusterization parameter
 $
 r_{\rm clustering}=\dfrac{\pi}{\hat m_{(-)} }\sim \dfrac{\pi}{
 H_0\Omega_R^{1/2} (1+z_r)} \sim 130\, {\rm Mpc}
 $
  recently
 discovered in studies of a large scale periodicity in redshift
 distribution \cite{a1}
 lead to a reasonable value of the radiation-type density
  $10^{-4}<\Omega_R\sim 3\cdot 10^{-3}<5\cdot 10^{-2}$ at the time of this
  epoch.


\section{Conclusions}

 The conservation of the number of variables of GR after the separation
 of the cosmological scale factor  from all fields
 leads to the Hamiltonian version of GR where the cosmic evolution
 acquires the pattern of a superfluid motion without the friction-type
 interaction.
  In this version the scale factor is an evolution parameter in
 the ``field space of events'' $[\vh|\widetilde{F}^{(n)}]$, and its
 canonical momentum (i.e. the Hubble ``parameter'') plays the role of the
 generator of evolution of the fields $\widetilde{F}^{(n)}$.
 The values of the scale factor momentum
  for solutions of the equations of motion can be called the
``reduced energies''.

  The solution of the problem of the ``negative reduce energy''  by the
 primary quantization and the
 secondary one (on the analogy of the pathway passed by QFT in
 the 20th century) reveals in GR other attributes of
 the theory of superfluid quantum liquid:
    London-type WDW  wave function
   and  Bogoliubov-type condensate of quantum universes.
 The postulate of the  quantum Bogoliubov vacuum as the state with
 the minimal ``energy'' leads to
   the absolute beginning of geometric time.

 The
 Hamiltonian approach
 leads to    potential perturbations of the scalar metric
 components in  contrast to the standard cosmological
 perturbation theory  \cite{Lif}  keeping only the kinetic
 perturbations  which are
 responsible for the ``primordial power spectrum'' in
 the inflationary model \cite{bard}.
The Quantum Gravity considered as the theory of superfluidity
gives us
 a  possibility  to explain this ``spectrum'' and
  other topical problems of cosmology
  by  the cosmological creation of the primordial W-, Z- bosons
  from vacuum, when
 their Compton length coincides with the universe horizon.

  The correspondence
   principle as the low-energy expansion of the reduced action
   identifies the conformal quantities with the ``measurable''  ones,
   and the uncertainty  principle  establishes the point of
   the beginning of the cosmological creation of the primordial W-, Z- bosons
  from vacuum due to their mass singularity at the moment
  $a_I^2\simeq 10^{-29}, H_I^{-1}\simeq 1$ mm. In this case,
 the equations describing the longitudinal
 vector bosons
 in SM  are close to the  equations
 of the inflationary model used for
 description of the ``power primordial spectrum'' of the CMB radiation
\cite{bard}.
 We listed the set of theoretical and observational
 arguments in favor of that the CMB radiation can be
 a final product of  primordial vector W-, Z- bosons cosmologically created
 from the vacuum.


{\bf Acknowledgement}\\
The authors are grateful to    A.A. Gusev, A.V. Efremov,
 E.A. Kuraev,   V.V. Nesterenko, V.B. Priezzhev,  and S.I. Vinitsky
 for interesting and critical
discussions. AFZ is grateful to the National Natural Science
Foundation of China (NNSFC) (Grant \# 10233050)  for a partial
financial support.

\end{document}